\pdfoutput=1
\documentclass[twocolumn,prd,superscriptaddress,preprintnumbers,nofootinbib]{revtex4}

\usepackage{graphicx}
\usepackage{epsfig}
\usepackage{lipsum}
\usepackage{enumitem}
\usepackage{bm}
\usepackage{latexsym,amssymb,amsmath,amsfonts,amssymb,txfonts,pxfonts,wasysym,float}
\usepackage{color}
\usepackage{mathrsfs}
\newcommand{\beq}[1]{\begin{equation}\label{#1}}
\newcommand{\eeq}{\end{equation}}
\newcommand{\bea}[1]{\begin{eqnarray} \label{#1}}
\newcommand{\eea}{\end{eqnarray}}
\newcommand{\ba}{\begin{array}}
\newcommand{\ea}{\end{array}}

\def\be{\begin{equation}}
\def\ee{\end{equation}}
\def\gs{\mathrel{
   \rlap{\raise 0.511ex \hbox{$>$}}{\lower 0.511ex \hbox{$\sim$}}}}
\def\ls{\mathrel{
   \rlap{\raise 0.511ex \hbox{$<$}}{\lower 0.511ex \hbox{$\sim$}}}}

\newcommand{\comment}[1]{}

\usepackage[usenames,dvipsnames]{xcolor}
\definecolor{orange}{cmyk}{0,0.5,1,0}
\definecolor{rossoCP3}{cmyk}{0,.88,.77,.40}
\definecolor{graa}{rgb}{0.8,0.8,0.8}
\definecolor{blaa}{rgb}{0.2,0.2,0.6}

\begin{document}

\title{\color{rossoCP3}{Hot thermal universe endowed with massive dark vector fields and the Hubble tension}}

\author{Luis A. Anchordoqui}

\affiliation{Department of Physics and Astronomy,  Lehman College, City University of
  New York, NY 10468, USA
}

\affiliation{Department of Physics,
 Graduate Center, City University
  of New York,  NY 10016, USA
}

\affiliation{Department of Astrophysics,
 American Museum of Natural History, NY
 10024, USA
}

\author{Santiago E. Perez Bergliaffa}

\affiliation{Department of Physics and Astronomy,  Lehman College, City University of
  New York, NY 10468, USA
}

\affiliation{Departamento de F\'{\i}sica Te\'orica, Instituto de F\'{\i}sica, Universidade do Estado de Rio de Janeiro, CEP 20550-013, Rio de Janeiro, Brasil}

\begin{abstract}
  \vskip 2mm \noindent
The value of the Hubble constant inferred from Planck measurements of anisotropies in the cosmic microwave background is at $4.4\sigma$ tension with direct astronomical measurements  at low redshifts. Very recently, it has been conjectured that this discrepancy may be reconciled if a small fraction of the dark matter is described by three  mutually orthogonal vector fields of the same mass. We study the thermal description of this model and use the
observationally-inferred primordial fractions of baryonic mass in
$^4{\rm He}$ to constrain its phase space.  We show that while the sterile
vector fields may help to alleviate a little bit the existing tension in
the measurements of the Hubble parameter, they cannot eliminate the
discrepancy between low- and high-redshift observations.
\end{abstract}
\maketitle


Over the past decade or so, cosmology has witnessed an avalanche of
data which has signaled the inception of ``precision cosmology.''
During this period and through many experiments, $\Lambda$CDM has
become established as a well-tested cosmological model. Within this
set up, the expansion of the universe today is dominated by  the cosmological constant $\Lambda$ and cold dark
matter (CDM). Nevertheless, various discrepancies have persisted. Most
strikingly, the emerging tension in the inferred values of the Hubble
constant
$H_0 = 100~h~{\rm km \, s^{-1} \,
  Mpc^{-1}}$~\cite{Freedman:2017yms}. $H_0$ parametrizes the expansion
rate and thus provides clues about the cosmological energy content of
the universe.

Numerous independent
measurements of $h$ at low-redshift, including those from Cepheids
and type-Ia supernovae, seem to indicate that $h = 0.735 \pm 0.016$~\cite{Riess:2011yx,Riess:2016jrr,Riess:2018byc,Bonvin:2016crt,Birrer:2018vtm}. All
these observations are local and consequently nearly independent of
the cosmological expansion history. On the other hand,  when the all-sky map
from the temperature
fluctuations on the cosmic microwave background (CMB) is combined with
data from Baryon Acoustic Oscillations (BAO) to calibrate the sound
horizon, the inferred $H_0$ value is $h = 0.6775 \pm
0.0075$~\cite{Hinshaw:2012aka,Ade:2015xua,Aghanim:2018eyx,Verde:2019ivm}. The
disagreement is significant at $4.4\sigma$
level~\cite{Verde:2019ivm,Riess:2019cxk}, and systematic effects do
not seem to be responsible for this
discrepancy~\cite{Follin:2017ljs,Dhawan:2017ywl,Shanks:2018rka,Riess:2018kzi,Bengaly:2018xko}. This then
substantiates a possible physics-beyond-$\Lambda$CDM origin of the
$H_0$ tension. For example, free-streaming relativistic particles from
a dark sector may alter the epoch of matter-radiation equality, boosting the expansion rate.

To accommodate new physics in the form of extra relativistic degrees
of freedom  it is convenient to account for the extra
contribution to the standard model (SM) energy density, by normalizing it to that of an
``equivalent'' neutrino species. The number of ``equivalent'' light
neutrino species,
\begin{equation}
N_{\rm eff} \equiv \frac{\rho_{\rm R} -
\rho_\gamma}{\rho_{\nu_L}} \,,
\label{neff}
\end{equation}
quantifies the total ``dark'' relativistic
energy density (including the three left-handedSM neutrinos) in units
of the density of a single Weyl neutrino:
\begin{equation}
\rho_{\nu_L} =\frac{7 \pi^2}{120} \
\left(\frac{4}{11}\right)^{4/3} \ T_\gamma^4 \,,
\end{equation}
where $\rho_\gamma$ is the energy density of
photons (with temperature $T_\gamma$) and $\rho_{\rm R}$ is the total
energy density in relativistic particles~\cite{Steigman:1977kc}. The
Hubble tension hints at the presence of an excess $\Delta N_{\rm eff}$
above the SM expectation. Note that the
normalization of $N_{\rm eff}$ is such that it gives $N_{\rm eff}^{\rm SM} = 3$ for three families of massless 
left-handed  neutrinos. For most practical purposes, it is
accurate enough to consider that SM neutrinos freeze-out completely at about
1~MeV. However, as the temperature dropped below this value, these neutrinos
were still interacting with the electromagnetic plasma and hence received a
tiny portion of the entropy from pair annihilations. The non-instantaneous
neutrino decoupling gives a minor correction to the SM contribution $\Delta
N_{\rm eff}^{\rm SM} = 0.046$~\cite{Mangano:2005cc}. 

The impact of
the $h$ determination is particularly complex in the investigation
of $N_{\rm eff}$. For example, combining CMB observations with BAO data the Planck Collaboration reported \mbox{$N_{\rm
  eff} = 2.99 \pm 0.17$~\cite{Aghanim:2018eyx}.} However, a combination
of the space telescope measurement $h = 0.738 \pm 0.024$ with the
Planck CMB data gives $N_{\rm eff} = 3.62 \pm 0.25$, which suggests
new neutrino-like physics (at around the $2.3\sigma$
level)~\cite{Ade:2015xua}. Finally, the simultaneous fit  to: {\it
  (i)}~CMB observations, {\it (ii)}~lensing and BAO data, and {\it (iii)}~local $H_0$ measurements leads to
$N_{\rm eff} = 3.27 \pm 0.15$ and $h = 0.6932 \pm
0.0097$~\cite{Aghanim:2018eyx}. In addition, light-element abundances probing big-bang nucleosynthesis
(BBN) have also hinted at the presence of extra relativistic degrees
of freedom. More concretely, the observationally-inferred primordial fractions
of baryonic mass in $^4{\rm He}$ favor $N_{\rm eff} =
3.80^{+0.80}_{-0.70}$ (with $2\sigma$ errors)~\cite{Izotov:2010ca}.

Proposed modifications of $\Lambda$CDM  to accommodate the $H_0$
tension reshape either the local expansion rate or else the early universe
pre-CMB emission. Among the plethora of proposals that have
been put forward are those based on sterile neutrinos~\cite{Anchordoqui:2011nh,Anchordoqui:2012qu,Jacques:2013xr,Brust:2013xpv,Gelmini:2019deq}, Goldstone
bosons~\cite{Weinberg:2013kea},  axions~\cite{Poulin:2018dzj,DEramo:2018vss}, scalar fields~\cite{Poulin:2018cxd,Agrawal:2019lmo,Lin:2019qug,Agrawal:2019dlm}, and decaying dark
matter~\cite{Menestrina:2011mz,Feng:2011in,Berezhiani:2015yta,Anchordoqui:2015lqa,Vattis:2019efj,Desai:2019pvs}. In a
similar fashion, it has been recently conjectured that the $H_0$ tension may be
reconciled if a small fraction of the dark matter is described by
three mutually orthogonal dark vector fields of the same
mass~\cite{Flambaum:2019cih}. In this paper we study the
thermal description of this model and use tests of BBN to constrain its phase space.  We show that while the sterile
vector fields may help to alleviate a little bit the existing tension in
the measurements of the Hubble parameter, they cannot eliminate the
discrepancy between low- and high-redshift observations.

The  $\Lambda$CDM extension under investigation evolves in a
Friedmann-Lem\^aitre-Robertson-Walker geometry, with
generic metric  
\begin{equation}
ds^2=-dt^2+a(t)^2(dr^2+r^2d\Omega^2) \, ,
\end{equation}
where the scale factor $a(t)$ is normalized such that at the present
time we have $a = 1$. The renormalizable Lagrangian density of the model can be expressed in
terms of the visible (SM) and dark sectors
\begin{equation}
  \mathscr{L} = \mathscr{L}_{\rm SM} +  \mathscr{L}_{\rm dark}  \,,
\end{equation}
with
\begin{equation}
\label{lagr}
\mathscr{L}_{\rm dark}  \supset \sqrt{-g}\left(-\frac 1 4 \mathbf{F}_{\mu\nu}
  \mathbf{F}^{\mu\nu} -\frac{m^2}{2}\mathbf{A}_\mu \mathbf{A}^{\mu}
\right)  \,,
\end{equation}
where $\mathbf A^\mu = (A^\mu_1,A^\mu_2,A^\mu_3)$ represents a set of
three massive vector fields and $\mathbf F_{\mu\nu} = \partial_\mu \mathbf A_\nu - \partial_\nu \mathbf A_\mu$. 
The equation of motion
that follows from the Lagrangian of the dark vector fields (\ref{lagr}) is found to be
\begin{equation}
\frac{1}{\sqrt{-g}}\partial_\mu(\sqrt{-g}\:\mathbf F^{\mu\nu} )  - m^2\mathbf A^{\nu}=0.
\end{equation}
Using the decomposition 
$\mathbf{A}^\mu 
= (- \boldsymbol{\phi}, \vec{\mathbf{A}} )  $, 
 the electric and magnetic fields can be defined as
\begin{equation}
\label{fields}
\vec{\mathbf{E}} = -\dot{\vec{\mathbf{ A}}}-\vec\nabla
\boldsymbol{\phi} \quad {\rm and} \quad
\vec{\mathbf B}=\vec \nabla \times \vec{\mathbf A},
\end{equation}
where a dot represents the derivative with respect to time. Now,
restricting to homogeneous solutions, it is straightforward to see that $\phi =0$ and
$\vec{\mathbf{A}}$ obeys
\begin{equation}
\label{eqA}
\ddot{\vec{\mathbf A}}+\frac{\dot a}{a} \ \dot{\vec{\mathbf
    A}}+m^2 \ {\vec{\mathbf A}}=0 \, ;
\end{equation}
see~\cite{Flambaum:2019cih} for details.
The solutions of (\ref{eqA}) yield, through \eqref{fields}, a time-dependent electric field with 
a null magnetic field.

The stress-energy tensor that follows from the Lagrangian density \eqref{lagr} is
\begin{equation}
\label{Tmunu}
T_{\mu\nu}= \mathbf{F_{\mu\alpha}}\mathbf{F^{\mu\alpha}}-
\frac 1 2 g_{\mu\nu}
\left(
\frac 1 2 \mathbf{F_{\alpha\beta}}\mathbf{F^{\alpha\beta}}
+
m^2\mathbf{A}_\alpha \mathbf{A}^\alpha \right)+m^2 \mathbf{A}_\mu
\mathbf{A}_\nu \, .
\end{equation}
This tensor must be diagonal to comply  with the symmetries of the metric. It can be taken to a diagonal form by averaging over the three fields, when 
the they are chosen to be mutually orthogonal,  and with the same mass and norm $|\mathbf{A}_\mu \mathbf{A}^\mu|$. In this case, 
\begin{equation}
\label{Tespacial}
\overline{T}_{ij}=\frac 1 6 \delta_{ij}(\mathbf{E}^2-m^2\mathbf{ A}^2),
\end{equation}
where the bar denotes the average over ``the triplet.''
Combining \eqref{Tmunu} and \eqref{Tespacial} we obtain an expression
for the energy density
\begin{equation}
\rho_A=\frac{1}{2a^2}(\mathbf{E}^2+m^2 \mathbf{ A}^2),
\end{equation}
and another one for the pressure
 \begin{equation}
p_A=\frac{1}{6a^2}(\mathbf{E}^2-m^2 \mathbf{ A}^2).
 \end{equation}
 Hence, the equation of state is
 \begin{equation}
 w \equiv \frac {p_A }{\rho_A} = \frac 1 3 \frac{\mathbf{E}^2-m^2 \mathbf{ A}^2}{\mathbf{E}^2+m^2 \mathbf{ A}^2},
\end{equation}
with $-1/3 < w < 1/3$.

In the radiation-dominated epoch, the scale factor can be written as 
\begin{equation}
a(t) = \left(2\sqrt{\Omega_{\rm R0}} \,H_0t \right)^{1/2} \, ,
\label{adet}
\end{equation}
where $\Omega_{\rm R}$ is the radiation density normalized to the
critical density and the subindex zero indicates it is evaluated
today. Using this expression, a particular solution of \eqref{eqA} can be written as follows:
\begin{equation}
\mathbf{ A}(t) = \mathbf{ A}_0 \ [mt]^{1/4} \ \left[c_1J_{1/4}(mt)+c_2Y_{1/4}(mt)    \right],
\end{equation}
where $J(x)$ and $Y(x)$ are Bessel functions, $\mathbf{ A}_0$
is a constant vector, and $c_1$ and $c_2$ are constants. With the
expression for $\mathbf{ A}(t)$, we can calculate the energy density, which has the following asymptotic limits:
\begin{equation}
\rho_A \approx |\mathbf{A_0}|^2
\left\{
\begin{array}{ll}
\cfrac{\sqrt 2 m (c_1+c_2)^2}{\Gamma^2(1/4)}\cfrac{1}{a^2t}\,, &  ~~{\rm{for}}\; mt\ll 1 \\
\\
2m^2(c_1^2+c_2^2)\cfrac{1}{a^2\sqrt{mt}} \,, & ~~{\rm{for}}\; mt\gg 1
\end{array}
\right. .
\label{rhoarray}
\end{equation}
Substituting (\ref{adet}) into (\ref{rhoarray}) it is easily seen that $\rho_A \propto a^{-4}$ for $t \ll m^{-1}$, and 
$\rho_A \propto a^{-3}$ for $t \gg m^{-1}$, so that the energy density of the set of vector fields, when conveniently averaged, behaves as radiation first and then as matter, always during the radiation phase. 

In the light of this behaviour, a modification to the
functional form of the $\Lambda$CDM Hubble parameter, 
\begin{equation}
E(a) =
\left\{\frac{\Omega_{R0}}{a^{4}}+\frac{\Omega_{m0}-\Omega_{A0}}{a^{3}}+\frac{\Omega_{A0}}
  {a^{3(w +1)}}+\Omega_\Lambda
\right\}^{1/2},
\label{Ea}
\end{equation}
was proposed in \cite{Flambaum:2019cih},
with $\Omega_{R0} = \Omega_{\gamma0} + \Omega_{\nu0}$, $\Omega_{\gamma
0} \simeq 5.37 \times 10^{-5}$, $\Omega_{\nu 0} \simeq 3.66 \times 10^{-5}$,
$\Omega_{m0} = \Omega_{b0} + \Omega_{\rm CDM0}$,
$\Omega_{b0} \simeq 0.048$, $\Omega_{\rm CDM0} \simeq 0.258$,
$\Omega_{\Lambda} \simeq 0.692$~\cite{Tanabashi:2018oca}. Here, $\Omega_{A0}$ is a free parameter
of the model with $w = 1/3$ for $mt<1$, and 
$w=0$ for $mt>1$. The extra contribution to the radiation density at early times leads to
an increment of the Hubble parameter
\begin{equation}
  H \equiv H_0 \ E(a) \,,
\label{Ha}
\end{equation}
 with respect to its value in the SM.

Next, in line with our stated plan, we use experimental data to
constrain the model.  BBN is the earliest observationally verified
landmark at a temperature of
$T_{\rm BBN} = (1+z) T_{\gamma 0} \simeq 2.348 \times 10^{-4}~{\rm
  GeV}$, where $T_{\gamma 0} = 2.348 \times 10^{-4}~{\rm eV}$ and the
redshift is $z \simeq 10^9$. Using the number of relativistic degrees of
freedom at temperature $T$,
\begin{equation}
N_{\rm R}(T)=\sum_{B}g_B \ \left(\frac{T_B}{T} \right)^4+\frac 7 8
\sum_{F} g_F \ \left(\frac{T_F}{T} \right)^4,
\label{ner}
\end{equation}
we can write the total energy density of all types of relativistic
particles in thermal equilibrium as
\begin{equation}
\rho_{\rm R} = \frac{\pi^2}{30}N_{\rm R}(T)T^4 \, ,
\end{equation}
where $g_{B(F)}$ is the total number of boson (fermion) degrees of
freedom and the sum runs over all boson (fermion) states with $m_i\ll
T$~\cite{Kolb:1990vq}. The factor of  $7/8$ is due to the difference between the Fermi and
Bose integrals. (\ref{ner}) defines the effective number of degrees of freedom,
$N_{\rm R}(T)$, by taking into account new particle degrees of freedom as the temperature is raised. Comparing the Friedmann equation,  
\begin{equation}
  H= 1.66 \ \sqrt{N_{\rm R} (T)}\;\frac{T^2}{M_{\rm Pl}}\,,
  \label{Friedmann}
\end{equation}
with (\ref{Ea}) and (\ref{Ha}), and using $a = (1+z)^{-1}$, we obtain a relation between $\Omega_{A0}$ and $N_{\rm
  R} (T)$. The number of equivalent light neutrino species now follows
from (\ref{neff}) and is given by
\begin{equation}
  N_{\rm eff} (T) = \frac{4}{7} \left[N_{\rm R} (T) -2\right]
  \left(\frac{11}{4}\right)^{4/3}  \, ,
\label{neff2}
\end{equation}
assuming that all neutrinos flavors decouple at the same
temperature. The $2\sigma$ upper limit
$\Delta N_{\rm eff} (T_{\rm BBN}) = 1.554$ constrains $\Omega_{A0}$
via (\ref{Ea}), (\ref{Ha}), (\ref{Friedmann}), and (\ref{neff2}). We note in passing that for $h = 0.678$ $\Lambda$CDM gives
$N_{\rm eff} (T_{\rm BBN}) \simeq 2.99$, whereas
for $h = 0.69$ $\Lambda$CDM predicts
$N_{\rm eff} (T_{\rm BBN}) \simeq 3.24$; both these results are in
agreement with those reported in~\cite{Aghanim:2018eyx}. For the
model at hand, we find that the region of the parameter space that
predicts $\Omega_{A0} \geq 5 \times 10^{-6}$ is excluded at the
$2\sigma$ level.

There is a second theoretical constraint on the model, which is more
restrictive. Since the dark vector fields must decouple before the
electroweak phase transition there is a reheating of the left-handed
neutrinos with respect to the dark vector fields. The reheating
temperature can be computed comparing the 106.75 degrees of freedom of
the SM with the 10.75 degrees of freedom of the primordial plasma
before neutrino decoupling~\cite{Kolb:1990vq}. This leads to
$T_A/T_\nu = (10.75/106.75)^{1/3}$. By taking into account that each
massive vector boson has 3 degrees of freedom and counting the 6
degrees of freedom of the left-handed neutrinos we obtain
$\Delta N_{\rm eff} (T_{\rm BBN}) = 0.24$. Since it is not possible to
reach $\Delta N_{\rm eff} (T_{\rm BBN}) = 0.24$ for a value
$h = 0.73$, we take instead $h= 0.69$. The latter requires 
$\Omega_{A0} = 2 \times 10^{-7}$. Finally, we use the 
constraint from the comoving
sound horizon to show that the mass of the vector field satisfies $3.6
\times 10^{-28} < m/{\rm eV} < 10^{-5}$ at the
$1\sigma$ level; see Appendix.

In summary, we have examined the hot thermal description of the model
proposed in~\cite{Flambaum:2019cih} to resolve the Hubble tension.
We have shown that while the addition of three dark massive
vector fields may help to alleviate a little bit the existing tension in
the measurements of the Hubble constant, it cannot eliminate the
discrepancy between low- and high-redshift observations.

\acknowledgments{LAA is supported by the by the U.S. National Science Foundation (NSF
    Grant PHY-1620661) and the National Aeronautics and Space
    Administration (NASA Grant 80NSSC18K0464).  S.E.P.B. is supported by
  Coordena\c c\~ao de Aperfei\c coamento de Pessoal de N\'{\i}vel
  Superior-Brasil (CAPES)-C\'odigo de Financiamento 001, and UERJ. He
 thanks Lehman College (CUNY) for hospitality. Any opinions,
    findings, and conclusions or recommendations expressed in this
    material are those of the authors and do not necessarily reflect
    the views of the NSF or NASA.}

\section*{Appendix}

There is abundant evidence supporting 
the existence of acoustic waves
that travelled through the tightly-coupled system of baryons and photons that 
filled the early universe~\cite{Aubourg:2014yra}. These waves propagated until the decoupling of the photons at $z=1100$, with a velocity given by
\begin{equation}
c(a) =
\frac{c}{\sqrt{3\left(1+\cfrac{3a\Omega_b}{4\Omega_\gamma}\right)}}
\, .
\end{equation}
The sonic horizon, {\it{i.e.}} the distance that a sound wave could have travelled since since $t=0$ up to the recombination time $t_*$ is given by
\begin{equation}
r_s = \int_0^{t_*} c_s dt \,, 
\end{equation}
which can be written as 
\begin{equation}
\frac{r_s H_0}{c} = \int_0^{a_*} \cfrac{c_s(a)/c}{a^2E(a)} \ da \equiv
P^{-1} \, .
\label{pm1}
\end{equation}
Observation leads to $P =
29.63^{+0.48}_{-0.45}$~\cite{Aubourg:2014yra}. Since $r_s$ can be
determined for any given cosmological model, the value of $P$ can be
used to obtain $H_0$, and a decrement of the value of $r_s$ leads to
an increment of $H_0$. Numerical integration of (\ref{pm1}) with $\Omega_{A0} = 2
\times 10^{-7}$ and $h = 0.69$ requires 
$3.6 \times 10^{-28} < m/{\rm eV} < 10^{-5}$ to satisfy the $P$ constraint at
$1\sigma$, and the lower limit from radiation-matter equivalence $m>
3.6 \times 10^{-28}~{\rm eV}$~\cite{Flambaum:2019cih}.

\end{document}